\documentclass{emulateapj}

\def\lea{\mathrel{<\kern-1.0em\lower0.9ex\hbox{$\sim$}}}
\def\gea{\mathrel{>\kern-1.0em\lower0.9ex\hbox{$\sim$}}}

\shorttitle{Connection between Star-forming, AGN, and Early-Type Galaxies} \shortauthors{Lee et al.}

\begin{document}

\title{THE CONNECTION BETWEEN STAR-FORMING GALAXIES, AGN HOST GALAXIES AND EARLY-TYPE GALAXIES IN THE SDSS} 

\author{Joon Hyeop Lee$^{1}$, Myung Gyoon Lee$^{1}$, Taehyun Kim$^{1}$, Ho Seong Hwang$^{1}$, Changbom Park$^{2}$, and Yun-Young Choi$^{2}$} 
\affil{$^{1}$ Astronomy Program, Department of Physics and Astronomy, Seoul National University, Seoul 151-742, Korea \\
    $^{2}$ Korea Institute for Advanced Study, Dongdaemun-gu, Seoul 103-722, Korea \\
	email: jhlee@astro.snu.ac.kr, mglee@astrog.snu.ac.kr, thkim@astro.snu.ac.kr, hshwang@astro.snu.ac.kr, cbp@kias.re.kr, yychoi@kias.re.kr }

\begin{abstract}

We present a study of the connection between star-forming galaxies, AGN host galaxies,
and normal early-type galaxies in the Sloan Digital Sky Survey (SDSS).
Using the SDSS DR5 and DR4plus data, we select our early-type galaxy sample in the color versus color-gradient space,
and we classify the spectral types of the selected early-type galaxies into normal, star-forming, Seyfert, and LINER classes, using several spectral line flux ratios.
We investigate the slope in the fundamental space for each class of early-type galaxies
and find that there are obvious differences in the slopes of the fundamental planes (FPs)
among the different classes of early-type galaxies, in the sense that the slopes for Seyferts and star-forming galaxies are flatter than those for normal galaxies and LINERs. 
This may be the first identification of the systematic variation of the FP slope among the subclasses of early-type galaxies.
The difference in the FP slope might be caused by the difference in the degree of nonhomology among different classes or by the difference of gas contents in their merging progenitors. One possible scenario is that the AGN host galaxies and star-forming galaxies are formed by gas-rich merging and that they may evolve into normal early-type galaxies after finishing their star formation or AGN activities.

\end{abstract}


\section{INTRODUCTION}

One of the interesting issues of modern observational cosmology is the evolutionary connection between various classes of galaxies: early-type galaxies, late-type galaxies, dwarf galaxies, mergers, star-forming galaxies, AGN host galaxies, and so on. Such connections, if found, may be important constraints on the galaxy formation and evolution scenario. 
Early-type galaxies are an ideal target to investigate this connection problem.
Early-type galaxies were considered for a long time to belong to a single family. However, recent studies found that there are variations of earl-type galaxies.
For example, some early-type galaxies have young stellar populations unlike typical normal early-type galaxies \citep{abr99,men01}, and some early-type galaxies have active nuclei (AGNs) in their centers \citep{san04,cap06}.
These abnormal classes of early-type galaxies are possibly the links between normal early-type galaxies and other classes of galaxies, 
and the relationship between normal early-type galaxies and other classes of galaxies may be important for constraining the formation model of early-type galaxies.
Recently, \citet{lee06} suggested that there are possible evolutionary connections between star-forming galaxies, AGN host galaxies, and normal early-type galaxies, from the study of ``blue early-type galaxies'' in the GOODS \citep[Great Observatories Origins Deep Survey;][]{gia04} {\it HST}/ACS fields.

The fundamental plane \citep[FP;][]{dre87,djo87} of galaxies provides important clues about the relationship among various classes of galaxies.
\citet{tac02} studied the properties of 18 ultraluminous infrared galaxies (ULIRGs) and concluded that those ULIRGs may evolve into normal elliptical galaxies based on their position in the fundamental space.
It is also known that AGN host galaxies reside at the same plane as the normal early-type galaxies in the fundamental space \citep{sne03, woo04}, which implies that those active galaxies and normal early-type galaxies have a very close relationship in the evolutionary sequence.
However, it was difficult to secure a large sample of abnormal classes of early-type galaxies (e.g., AGN host early-type galaxies or post--star-forming early-type galaxies) in the previous studies.

In this Letter, we present a study of the connection between the various sub-classes of
early-type galaxies using the large sample in the Sloan Digital Sky Survey \citep[SDSS;][]{yor00}.
The outline of this Letter is as follows. 
Section 2 describes the data set we used and the methods to select early-type galaxies and to classify early-type galaxies into several sub-classes.
We present the FP analysis of each class of early-type galaxies in \S3, and, finally, discussion and conclusion are given in \S4.
Throughout this Letter, we adopt the cosmological parameters 
$h=0.7$, $\Omega_{\Lambda}=0.7$, and $\Omega_{M}=0.3$.

\section{DATA AND GALAXY CLASSIFICATION}

We used the SDSS Data Release 5 \citep[DR5;][]{ade07} in this study. The DR5 imaging data cover about 8000 deg$^{2}$ in the $ugriz$ bands, and the DR5 spectroscopic data cover 5600 deg$^{2}$ \footnote{See http://www.sdss.org/dr5/.}. The photometric and spectroscopic observations were conducted with the 2.5 m SDSS telescope at the Apache Point Observatory in New Mexico between 1999 March and 2005 June.

\begin{figure}
\plotone{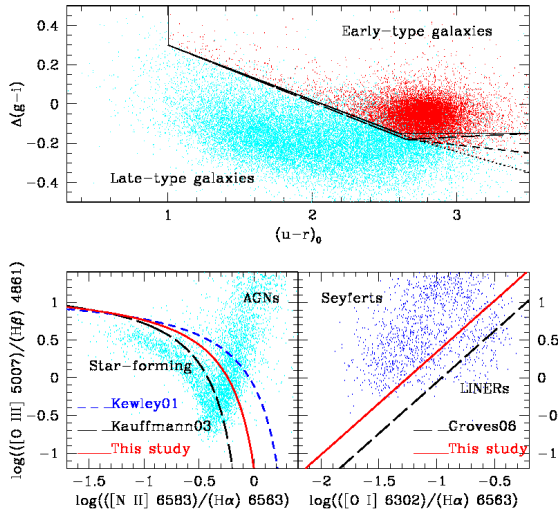}
\caption{
{\it Upper panel}: Segregation between early-type galaxies (dark gray points) and late-type galaxies (light gray points) in the color vs. color-gradient space.
The color gradient $\Delta (g-i)$ is defined as the difference in $(g-i)$ color between the region at $R<0.5R_{\textrm{\footnotesize pet}}$ and the region at $0.5R_{\textrm{\footnotesize pet}}<R<R_{\textrm{\footnotesize pet}}$ [negative $\Delta (g-i)$ for blue outside].
The lines represent different segregation guidelines for different magnitude ranges (solid line for $14.5<r_{pet}<16.0$, short-dashed line for $16.0<r_{pet}<16.5$, long-dashed line for $16.5<r_{pet}<17.0$, and dotted line for $17.0<r_{pet}<17.5$). 
{\it Lower left panel}: AGN selection in the BPT diagram \citep{bal81}. 
The short-dashed line is the theoretical upper limit of star-forming galaxies \citep{kew01}, and the long-dashed line is the empirical criterion of \citet{kau03}.
In this study, we used our empirical criterion ({\it solid line}) to distinguish AGNs from star-forming galaxies.
{\it Lower right panel}: Seyfert-LINER segregation in the [O III]/[H$\beta$] vs. [O I]/[H$\alpha$] diagram.
The solid line is our criterion, and the long-dashed line is that of \citet{gro06}.
\label{fig1}}
\end{figure}

We used only the galaxy catalog with spectroscopic information, which includes 573,113 available objects. 
From this catalog, we selected early-type galaxies with the galaxy classification method using the color versus color-gradient space \citep{par05} as shown in Fig. 1. 
In this classification method, colors and color gradients are the main criteria for classification, and the inverse concentration (C$^{-1} \equiv$ R$_{50\%}/$R$_{90\%}$) cut is also adopted differentially for different magnitude ranges: C$^{-1}<0.43$ for $14.5<r_{Pet}<16.0$, C$^{-1}<0.45$ for $16.0<r_{Pet}<16.5$, C$^{-1}<0.47$ for $16.5<r_{Pet}<17.0$, and C$^{-1}<0.48$ for $17.0<r_{Pet}<17.5$.\footnote{R$_{n\%}$ is the $n\%$ Petrosian radius, and $r_{Pet}$ is the Petrosian magnitude in the $r$ band.}
\citet{par05} estimated the completeness and reliability of this classification method to be as large as $\sim88\%$ at $r_{Pet}<17.5$. The color-gradient estimation was conducted for the 389,789 objects in the SDSS DR4plus \citep{cho07} sample, which is one of the products of the New York University Value-Added Galaxy Catalog \citep{bla05}. 
Finally, we selected 139,183 early-type galaxies.

We selected AGN host galaxies in our early-type galaxy sample, using the line flux ratio diagram of 
[O III]/[H$\beta$] versus [N II]/[H$\alpha$] \citep[BPT diagram;][]{bal81} as shown in Fig. 1.
\citet{kew01} suggested a theoretical upper limit of star-forming galaxies in the BPT diagram, and
\citet{kau03} adopted their empirical criterion to select AGNs as drawn in Fig. 1.
However, since our sample shows small offsets from the previous samples in the BPT diagram, we used our own empirical criterion to segregate AGN host galaxies from star-forming galaxies:
[O III]$/$[H$\beta$] $=0.61/$([N II]$/$[H$\alpha$] $-0.25)+1.25$, which is intermediate between the criteria of \citet{kew01} and the criteria of \citet{kau03}.
Among the selected AGN host early-type galaxies, we distinguished Seyferts from LINERs 
in the [O III]/[H$\beta$] versus [O I]/[H$\alpha$] diagram.
Since there is also a small offset\footnote{We used the emission line fluxes measured by the SDSS pipeline \citep{sto02}, while \citet{kau03} and \citet{gro06} used those measured by the pipeline of \citet{tre04}.} between the sample of \citet{gro06} and our sample in this diagram as shown in Fig. 1, we used our own empirical guideline: [O III]/[H$\beta$] $=1.36$ [O I]/[H$\alpha$] $+1.7$.
This process returns 1913 star-forming galaxies, 1129 Seyfert galaxies, and 320 LINERs with a signal-to-noise ratio (S/N) of $>3$ for each line, in our early-type galaxy sample.

We corrected several quantities provided by the SDSS DR5, for this study.
The effective radius of each galaxy was corrected for its axis ratio and redshift, and 
the surface brightness was calculated using the effective radius corrected for
the redshift effect and for the cosmological dimming \citep{ber03a}. 
The velocity dispersion data were produced using an automated spectroscopic pipeline named IDLSPEC2D version 5 in the SDSS DR4plus (Schelegel et al. 2007, in preparation). We corrected the velocity dispersions for the aperture effect, adopting the method of \citet{jor95}.
For a reliable estimation of the FP, we used the galaxies with S/N$>10$ for their velocity dispersion, which returns 130,147 normal\footnote{We define \emph{normal} early-type galaxies as early-type galaxies without evidence of obvious line emission.} early-type galaxies, 1608 star-forming galaxies, 1050 Seyfert galaxies, and 293 LINER galaxies.

\section{THE FUNDAMENTAL PLANE}

We estimated the slope in the fundamental space for each class of early-type galaxies, using the ordinary least-square bisector method \citep{iso90}.
The normal early-type galaxies in the SDSS show a good FP relation, as already shown by \citet{ber03b}.
In addition, AGN host galaxies and a large fraction of star-forming galaxies also reside in the FP of normal early-type galaxies.
The slopes of normal early-type galaxies and LINER early-type galaxies are consistent within $1\: \sigma_{\textrm{\footnotesize slope}}$ error ($1.301\pm0.002$ for normal and $1.254\pm0.052$ for LINER).
However, the slopes of Seyfert galaxies and star-forming galaxies are significantly ($>8\: \sigma_{\textrm{\footnotesize slope}}$) smaller than those of normal and LINER galaxies ($1.113\pm0.023$ for Seyferts and $0.996\pm0.019$ for star-forming galaxies).

\begin{figure}
\plotone{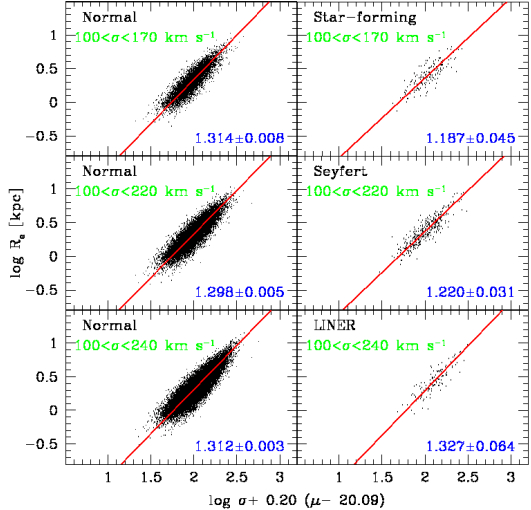}
\caption{
FP slope estimations with the same $\sigma_{v}$ distribution in given $\sigma_{v}$ ranges, for $0.04<z<0.1$.
{\it Upper panels}: Normal and star-forming early-type galaxies, with 100 km s$^{-1}$ $< \sigma_{v} <$ 170 km s$^{-1}$.
{\it Middle panels}: Normal and Seyfert early-type galaxies, with 100 km s$^{-1}$ $< \sigma_{v} <$ 220 km s$^{-1}$.
{\it Lower panels}: Normal and LINER early-type galaxies, with 100 km s$^{-1}$ $< \sigma_{v} <$ 240 km s$^{-1}$.
The lines are OLS bisector fittings, and the resulting FP slope and $1\: \sigma$ slope error are shown in the lower right-hand corner of each panel. Each velocity dispersion boundary was selected within the common range of the distributions of the two classes.
\label{fig2}}
\end{figure}

These differences in the FP slope, however, are possibly artifacts caused by the different parameter distributions of different classes in the fundamental space.
Since the slope of the FP varies in the different ranges of fundamental parameters \citep{ber03b}, it is fair to compare the FP slope with the galaxies in the same domain of fundamental space.
Among the three fundamental parameters, we set boundaries with velocity dispersion ($\sigma_{v}$), because this parameter may be less sensitive to galaxy evolution effects than effective radius or surface brightness.
The approximate range of $\sigma_{v}$ in each class is as follows: $20-420$ km s$^{-1}$ for normal galaxies, $20-170$ km s$^{-1}$ for star-forming galaxies, $20-220$ km s$^{-1}$ for Seyfert galaxies, and $100-240$ km s$^{-1}$ for LINER galaxies. 
Because the estimation error of $\sigma_{v}$ is very large for $\sigma_{v} < 100$ km s$^{-1}$ \citep{cho07}, we use galaxies with $\sigma_{v} > 100$ km s$^{-1}$ only.
In addition, it is safe to compare the galaxies at $0.04<z<0.1$, to minimize the selection effects caused by the fixed fiber-size and the LINER incompleteness \citep{kew06}.
Fig. 2 shows the FP slope estimates with the limits of $\sigma_{v}$ and redshift. The normal early-type galaxies were randomly resampled to have the same $\sigma_{v}$ distribution as the compared abnormal early-type galaxies. In this comparison, the difference between star-forming galaxies and normal galaxies is $2.8 \sigma_{slope}$, and the difference between Seyfert galaxies and normal galaxies is $2.5 \sigma_{slope}$, while LINER galaxies and normal galaxies have consistent FP slopes. Interestingly, Figure 4 of \citet{sne03} and Figure 3 of \citet{woo04} show that AGN host galaxies reside in the FP of normal early-type galaxies but that they are slightly tilted from the FP of normal early-type galaxies, for the same direction as our Seyfert galaxies are.

\begin{deluxetable}{cccc}
\tablecaption{Random Sampling Tests (N$_{\textrm{\footnotesize trial}}$=10000) with the Same Sample Size and $\sigma_{v}$ Distribution\label{tbl-1}} \tablewidth{0pt}
\tablehead{$\sigma_{v}$ Range & FP Slope of Class 1 & FP Slope &\\
(km s$^{-1}$) & For Normal Galaxies & of Class 2 & Error }
\startdata
$100-170$ & 1.315 & 1.187 \tablenotemark{a} & 0.051 \\
$100-220$  & 1.302 & 1.220 \tablenotemark{b} & 0.033 \\
$100-240$  & 1.302 & 1.327 \tablenotemark{c} & 0.054 \\
\enddata
\tablenotetext{a}{For star-forming galaxies.}
\tablenotetext{b}{For Seyfert galaxies.}
\tablenotetext{c}{For LINER galaxies.}
\end{deluxetable}

However, it is necessary to estimate the sampling error for each abnormal class of galaxies, because the sample sizes of abnormal galaxies are much smaller than that of normal galaxies. We conducted random sampling tests using the normal galaxy sample, with the same sample size and $\sigma_{v}$ distribution as each abnormal class. The random sampling was repeated 10,000 times for each comparison, and the results are summarized in Table 1. The error values in Table 1 include both estimation errors and sampling errors. Considering these errors, star-forming galaxies and Seyfert galaxies have different FP slopes from normal early-type galaxies with probabilities more than $98\%$ ($2.5\: \sigma$) in the same $\sigma_{v}$ and redshift ranges as the Fig. 2, and LINER galaxies and normal galaxies have consistent FP slopes.

One observational bias may affect the FP properties of the AGN host galaxies. Since spectroscopy was targeted to the central parts of galaxies, the $\sigma_{v}$ estimates of AGN host galaxies are possibly distorted by the existence of an AGN. However, it is difficult not only to correct such an effect but also to guess how much that changes the FP slope of AGN host galaxies, using current data.

\section{DISCUSSION AND CONCLUSION}

The FP of early-type galaxies is considered to be closely related to the virial theorem.
However, the virial plane (VP) does not match the FP exactly, and several studies have been conducted to investigate the origin of the tilt between the two planes.
\citet{bus97} found from the analysis of 40 nearby elliptical galaxies that the primary origin of the tilt between the VP and the FP is the non-homology of elliptical galaxies; that is, the velocity dispersion profiles of more massive galaxies are steeper than those of less massive galaxies. In addition to the non-homology effect, they also suggested that the effect of stellar populations is about $30\%$ responsible for the FP tilt. This result was supported by \citet{tru04}, who investigated the FP of 911 early-type galaxies in the combined catalog of the SDSS and 2MASS \citep[Two Micron All Sky Survey;][]{bel03}.

Adopting the nonhomology effect to the FP slope difference among different classes, it is inferred that the Seyfert and star-forming early-type galaxies may have stronger non-homology than the normal early-type galaxies.
It is not easy to tell what causes such strong non-homology in Seyfert and star-forming early-type galaxies, but it is possible to assume that some events causing their star formation or AGN activity might also affect (or be affected by) the dynamical structures of those classes.

Recently, several numerical experiments were conducted to explain the physical origin of the FP tilt.
Some self-consistent hydrodynamic simulations showed that the major merging with gas dissipation can reproduce the tilt between the VP and the FP \citep{ono05,dek06}.
\citet{rob06} studied the variation of the slope of the FP in diverse merging cases. They found that the dissipational merging produces the FP tilt, whereas the FP of the dissipationless merging is parallel to the VP. Furthermore, they showed that the merging of disk galaxies with a larger fraction of gas makes a larger tilt between the VP and the FP.

These previous results are very interesting, since they provide a hint to understanding the origin of the difference in the FP slopes of different classes of early-type galaxies.
According to those simulations, the different classes of our early-type galaxies might evolve from mergers with different gas contents.
Normal and LINER early-type galaxies might be formed by the merging of gas-poor disk galaxies, while Seyfert and star-forming early-type galaxies might be formed by the merging of gas-rich disk galaxies.
This interpretation is also consistent with the conclusion of \citet{kew06} that the major difference between Seyferts and LINERs may be the gas accretion rate.
However, it is expected that the star formation or AGN activities in abnormal early-type galaxies will not continue forever.
Since the structural parameters and the loci in the fundamental space of the abnormal classes are very similar to those of the normal early-type class, it is not too absurd to infer that the AGN host or star-forming early-type galaxies may evolve into normal early-type galaxies after finishing their star formation or AGN activities.

The observational analysis of \citet{kew03} led to an evolutionary scenario: starbursts initially triggered by tidal interactions $\rightarrow$ AGN activated by gas funneled toward the merger nucleus $\rightarrow$ circumnuclear star formation at late stages of the merger.
When combining this scenario with the hierarchical merging scenario of early-type galaxies \citep{too77,sea78}, it is possible to extend the scenario of \citet{kew03} as follows: mergers $\rightarrow$ galaxies with star formation or AGN $\rightarrow$ galaxies finishing their star formation or AGN activity $\rightarrow$ normal early-type galaxies. 
Considering that merging events can affect the FP slope, our results are consistent with such an evolutionary sequence, although our results do not prove that all early-type galaxies evolve along this sequence.

If the FP slope of the Seyfert and star-forming early-type galaxies become the same as that of normal early-type galaxies when their star formation or AGN activities finish, what physical mechanism works on that process? We do not have a clear answer to that question. Such a mechanism may need to explain the differential fading of galaxies with different size or mass, which can change the tilted FP slope of abnormal early-type galaxies into the same as that of normal early-type galaxies.

In this Letter, our conclusions are as follows:

1. We have found obvious differences in the FP slopes for different subclasses of early-type galaxies, from the analysis of the early-type galaxies in the SDSS.
Star-forming and Seyfert early-type galaxies show significant differences in their FP slope from normal early-type galaxies, while LINER early-type galaxies have a FP slope consistent with normal early-type galaxies.
This may be the first identification of the systematic difference in the FP slopes among the various subclasses of early-type galaxies.

2. The difference in the FP slope might be caused by the difference in the degree of nonhomology in different classes or by the difference of gas contents in their merging progenitors. One possible scenario is that the AGN host galaxies and star-forming galaxies are formed by gas-rich merging and that they may evolve into normal early-type galaxies after finishing their star formation or AGN activities.

\acknowledgments

This work was supported in part by a grant (R01-2004-000-10490-0) from the Basic Research Program of the Korea Science and Engineering Foundation.
CBP acknowledges the support of the Korea Science and Engineering
Foundation (KOSEF) through the Astrophysical Research Center for the
Structure and Evolution of the Cosmos (ARCSEC).
Funding for the SDSS and SDSS-II has been provided by the Alfred P. Sloan Foundation, 
the Participating Institutions, the National Science Foundation, 
the US Department of Energy, the National Aeronautics and Space Administration, 
the Japanese Monbukagakusho, the Max Planck Society, and the Higher Education Funding Council for England.
The SDSS Web site is http://www.sdss.org/.
The SDSS is managed by the Astrophysical Research Consortium for the Participating Institutions. 
The Participating Institutions are the American Museum of Natural History, 
Astrophysical Institute Potsdam, the University of Basel, the University of Cambridge, 
Case Western Reserve University, the University of Chicago, Drexel University, Fermilab, 
the Institute for Advanced Study, the Japan Participation Group, Johns Hopkins University, 
the Joint Institute for Nuclear Astrophysics, the Kavli Institute for Particle Astrophysics and Cosmology, 
the Korean Scientist Group, the Chinese Academy of Sciences (LAMOST), Los Alamos National Laboratory, 
the Max-Planck-Institute for Astronomy (MPIA), the Max Planck Institute for Astrophysics (MPA), 
New Mexico State University, Ohio State University, the University of Pittsburgh, the University of Portsmouth, 
Princeton University, the US Naval Observatory, and the University of Washington.

\end{document}